\documentstyle[psfig,aps]{revtex}
\newcommand{\bra}{\langle}
\newcommand{\ket}{\rangle}
\tightenlines
\begin{document}
\draft
\title{Chiral Baryon in the Coherent Pair Approximation}

\author{T. S. T. Aly}
\address{Faculty of Mathematics, Helwan University, Helwan, Egypt}

\author{J. A. McNeil}
\address{Physics Department, Colorado School of Mines, Golden,  Colorado 80401}

\author{S. Pruess}
\address{Department of Mathematical and Computer Sciences, Colorado School of Mines, Golden, Colorado 80401}

\maketitle
\begin{abstract}
We revisit the work of K. Goeke, et al. \cite{goeke} who considered a chiral model for the nucleon based on the linear sigma model with scalar-isoscalar scalar-isovector mesons coupled to quarks and solved using the coherent-pair approximation.   In this way the quantum pion field can be treated in a non-perturbative fashion.  In this work we review this model and the coherent pair approximation correcting several errors in the earlier work.  We minimize the expectation value of the chiral hamiltonian in the ansatz coherent-pair ground state configuration and solve the resulting equations for nucleon quantum numbers. We calculate the canonical set of nucleon observables and compare with the Hedgehog model and experiment.  Using the corrected equations yield slightly different values for nucleon observables but do not correct the large virial deviation in the $\pi$-nucleon coupling. Our results therefore do not significantly alter the conclusions of Goeke, et al..
\end{abstract}

\pacs{21.40.Aa}

\section{Introduction}

It is widely believed that quantum chromodynamics (QCD) is the fundamental theory underlying the strong interaction.  Regrettably, reliable first principle calculations of hadronic structure and reactions based on QCD are still some time off. Nevertheless, simpler QCD-motivated phenomenological models have been proposed which preserve the important property of chiral symmetry. Familiar examples include Skyrme-Witten models\cite{skyrme} and hybrid chiral-soliton models such as those of Kahana and Ripka \cite{kahana}, Birse and Banerjee\cite{bb}, Birse \cite{birse}, Fiolhais, Goeke, Gr\"ummer, and Urbano \cite{fiolhais}, Goeke, Harvey, Gr\"ummer, and Urbano \cite{goeke} hereafter referred to as GHGU and others (see Ref.\cite{christov}).  

Such  approaches, in particular Ref.\cite{bb}, argue that spontaneous symmetry breaking of the QCD Lagrangian gives rise to an effective chiral Lagrangian of the Gell-Mann-L\'evy sigma-model form involving explicit quark, scalar-isoscalar meson (sigma, $\sigma$), and pseudoscalar-isovector meson (pion, $\vec\pi$) degrees of freedom. The model has been solved in mean-field using the ``Hedgehog'' ansatz which assumes a configuration-space-isospin correlation for the pion field, $\vec\pi = \hat r \pi$, and for the quarks. One drawback to this ansatz is that it breaks both rotational and isospin invariance (although ``grand spin'', $\vec K =\vec I +\vec J$, remains conserved) requiring some projection onto physical states at the end. Considerable attention has been  given to the problem of projection in the calculation of observables \cite{cohen}.

In spite of this drawback the model is quite successful at predicting baryon properties.  Constraining the pion mass and decay constant with experimental values, the model contains but two additional free parameters (which can be written in terms of the effective quark and sigma masses), yet the model makes quite respectable  predictions for a host of hadronic properties (mass, magnetic moments, sigma commutator, pion-nucleon coupling constant ($g_{\pi NN}$), axial vector coupling constant ($g_A$), as well as weak and electromagnetic form factors).

Another criticism of this approach, however, is the approximate treatment of the pion whose very light mass argues for a quantum, as opposed to mean-field, treatment.  However, treating light strongly-coupled fields is in general difficult as perturbative methods are unreliable. In part to address such issues, the coherent-pair approximation was developed by Bolsterli \cite{bolsterli} and used by GHGU to treat the pions as true quantum fields non-perturbatively.  We find this approach promising and in this work  revisit the hybrid chiral model of Goeke, et al. correcting several errors.  Besides the inclusion of quantum pionic degrees of freedom non-perturbatively, another advantage is that one need not invoke the so-called Hedgehog ansatz for the pion field.  Specifically, the symmetry of the quarks with respect to permutation induces the space-isospin correlations in the pion field.  Following the ground-breaking work of GHGU, we minimize the chiral hamiltonian with respect to the coherent-pair Fock-space ansatz ground-state configuration, then calculate nucleon properties, and compare with other chiral models. We find that once minimization with respect to the coherence parameter is achieved, the results for calculated observables are not significantly different from those calculated in GHGU. In addition we present results for the sigma commutator which measures the degree of chiral symmetry breaking. 

For completeness and ease of comparison the organization of this paper closely follows that of GHGU.  We present the starting Lagrangian and the major intermediate results, refer the reader to the original paper for details, and note where our results differ from GHGU.  Where appropriate, we will give the corresponding GHGU equation references. We present the background theory in section II, a summary of the coherent-pair approximation results in section III, the variational equations in section IV, the derived nucleon properties in section V,  the numerical results in section VI, and our summary and conclusions in section VII.  We find that the corrected equations do not significantly alter the results originally reported in GHGU.  When compared  with the Hedgehog model the coherent-pair approximation shows systematically smaller mesonic contributions to the nucleon observables and the energy densities which may be related to poor self-consistency in the pionic sector.

\section{Chiral Quark-Meson Model}

We begin with the chiral model Lagrangian of Gell-Mann and L\'evy \cite{gellmannlevy} but with explicit quark degrees of freedom.  (Discussion of how such a form may be argued from QCD are given in Refs. [2 and 5].) After chiral symmetry breaking, inducing a pion mass, the Lagrangian can be written (GHGU Eq.[2.1]) as
\begin{equation}
{\cal L}(x)=i\hat{\bar{\psi}} \partial \hspace{-.08in}/ \hat{\psi}+\frac{1}{2}(\partial_{\mu}\hat{\sigma} \partial^{\mu}\hat{\sigma}+\partial_{\mu}\hat{\vec{\pi}}\cdot\partial^{\mu}\hat{\vec{\pi}})+g\hat{\bar{\psi}}(\hat{\sigma}+i\gamma_{5}\vec{\tau}\cdot\vec{\pi})\hat{\psi}-{\cal U}(\hat{\sigma},\hat{\vec{\pi}}),
\end{equation}
with
\begin{equation}
{\cal U}(\hat{\sigma},\hat{\pi})=\frac{\lambda^{2}}{4}(\hat{\sigma}^{2}+\hat{\vec{\pi}}^{2}-\nu^{2})^{2}-f_{\pi}m^{2}_{\pi}\hat{\sigma},
\end{equation}
where the carat ($\hat{~}$) denotes a quantum field, $f_{\pi}$ is the pion decay constant, $m_{\pi}$ is the pion mass, and $\nu$, $g$ and $\lambda$ are constants to be determined. In the standard scenario spontaneous symmetry breaking generates masses for the quark and sigma fields and the linear sigma term, which breaks the chiral symmetry and generates the small pion mass which would be zero otherwise as the Goldstone boson of the theory. The vacuum then has a unique non-vanishing scalar field expectation value,
\begin{equation}
{\partial U\over \partial\hat{\vec\pi}}=0\Longrightarrow   \vec\pi_0=0, ~~~~~~~~~~~~{\partial U\over  \partial\hat{\sigma}}=0\Longrightarrow  \sigma_0=f_{\pi}.
\end{equation}
Then the three undetermined constants in the original Lagrangian can be written in terms of the three effective masses: $m_q=-g\sigma_0$, $m_{\sigma}^2=\lambda^2(3 \sigma_0^2-\nu^2)$, and $m_{\pi}^2=\lambda^2(\sigma_0^2-\nu^2)$. 

We take the experimental values $f_{\pi}=93$ MeV and $m_{\pi}=139.6$ MeV, leaving $g$ and $m_{\sigma}$ as the only free parameters which must be determined.  The additional parameters introduced by the coherent-pair approximation are constrained by minimization.

Introducing the conjugate momenta, one formally converts the Lagrangian density to a Hamiltonian density (GHGU Eq. [2.10]) into
\begin{equation}
\begin{array}{l}
\hat{H}(r)=\frac{1}{2}\{ \hat{P}_{\sigma}(r)^{2}+[\nabla\hat{\sigma}(r)]^{2}+\hat{\vec{P}}_{\pi}(r)^{2}+[\nabla\hat{\vec{\pi}}(r)]^{2}\}\\ \hspace{.5in} +{\cal U}(\hat{\sigma},\hat{\vec{\pi}})+\hat{\psi}^{\dagger}(r)(-i\alpha  \vec{\nabla})\hat{\psi}(r)\\ \hspace{.5in} -g\hat{\psi}^{\dagger}(r)(\beta\hat{\sigma}(r)+i\beta\gamma_{5}\vec{\tau}\cdot\hat{\vec{\pi}})\hat{\psi}(r),
\end{array}
\end{equation}
where $\vec\alpha$ and $\beta$ are the usual Dirac matrices. In the above expression $\hat{\psi},\hat{\sigma}$, and $\hat{\vec{\pi}}$ are quantized field operators with the appropriate static angular momentum expansions (GHGU Eqs. [2.11],  [2.17], and [2.13]),
\begin{equation}
\hat{\sigma}(r)=\int \frac{d^{3}k}{[(2\pi)^{3}2 \omega_{\sigma}(k)]^{\frac{1}{2}}}[\hat{c}^{\dagger}(k)e^{-ik\cdot r}+\hat{c}(k)e^{+ik\cdot r}],
\end{equation}
\begin{eqnarray}
\hat{\pi}(r)=[\frac{2}{\pi}]^{\frac{1}{2}}\int\limits_{0}^{\infty} dkk^{2}[\frac{1}{2\omega_{\pi}(k)}]^{\frac{1}{2}} \sum\limits_{\ell m w}& {\rm j}_{\ell} (kr){\rm Y}_{\ell  m}^{*}(\Omega_{r}) [\hat{a}^{ 1 w~\dagger}_{\ell m}(k) \nonumber\\ 
&+(-)^{m+w} \hat{a}^{1 -w}_{\ell -m}(k)],  \label{pionfield}
\end{eqnarray}
\begin{equation}
\hat{\psi}(r)=\sum\limits_{njmw}(\bra r\mid njmw\ket\hat{d}^{{1\over 2} w}_{njm}+\bra r\mid \overline{n}jmw \ket\hat{d}^{ {1\over 2} w~\dagger}_{njm}),
\end{equation}
where the $|njmw\ket$ and $|\overline{n} jmw\ket$ form a complete set of quark and antiquark spinors with angular momentum quantum numbers and spin-isospin quantum numbers j, m, and w, respectively. The notation is slightly altered from GHGU in that isospin labels appear as superscripts and spin labels appear as subscripts.  The corresponding conjugate momentum fields have the expansions (GHGU Eqs. [2.14] and [2.18]), 
\begin{equation}
\hat{P}_{\sigma}(r)=i \int_{0}^{\infty} d^{3}k[\frac{\omega_{\sigma}(k)}{2(2\pi)^{3}}]^{\frac{1}{2}} 
[\hat{c}^{\dagger}(k)e^{-i\vec k\cdot \vec r}-\hat{c}(k)e^{+i\vec k\cdot \vec r}],
\end{equation}
\begin{eqnarray} 
\hat{{P}_{\pi}}(r)=i [\frac{2}{\pi}]^{\frac{1}{2}}\int\limits_{0}^{\infty} dk k^{2}[\frac{\omega_{\pi}(k)}{2}]^{\frac{1}{2}} \sum\limits_{\ell m w}&{\rm j}_{\ell}(kr) {\rm Y}_{\ell m}^{*}  [\hat{a}^{ 1 w~\dagger}_{\ell m}(k) \nonumber\\ 
& - (-)^{m+w}\hat{a}^{1 -w}_{\ell -m}(k)]. \label{pionmom}
\end{eqnarray}
Here the $\hat{c}(k)$ destroys a $\sigma$-quantum with momentum $\vec{k}$ and frequency $\omega_{\sigma}(k)=(k^{2}+m_{\sigma}^{2})^{\frac{1}{2}}$ and $\hat{a}^{1 w}_{t\ell m}(k)$ destroys a pion with momentum $\vec{k}$ and corresponding $\omega_{\pi}=(k^{2}+m_{\pi}^{2})^{\frac{1}{2}}$ in isospin-angular momentum state $\{\ell m;t w\}$.    For convenience one constructs the configuration space pion field functions needed for the subsequent variational treatment by defining the alternative basis operators,
\begin{equation}
\hat b^{1 w}_{\ell m} = \int dk \\ k^2 \xi_{\ell}(k) \hat a^{1 w}_{\ell m}(k),
\end{equation}
where $\xi_{\ell}(k)$ is the variational function. Taking this over to configuration space defines the pion field function (GHGU Eq. [3.11]),
\begin{equation}
\phi_{\ell}(r)=\frac{1}{2\pi}\int dk k^2 \frac{\xi_{\ell}(k)}{[\omega_{\pi}(k)]^{1/2}} {\rm j}_{\ell}(r).
\end{equation}
In the following only the $\ell =1$ value is used and the angular momentum label will be dropped.

\section{Ground State Configuration Ansatz}
 
The ansatz Fock state for the nucleon is taken to be (GHGU Eq. [4.23])
\begin{equation}
|NT_{3}J_{z}\ket=[\alpha (| n\ket \otimes|P^0_0\ket)_{T_{3}J_{z}}+\beta(|n\ket \otimes |P^1_1\ket)_{T_{3}J_{z}}+ \gamma(|\delta >\otimes |P^1_1\ket)_{T_{3}J_{z}}]|\Sigma \ket,
\end{equation}
where $|\Sigma \ket$ is the coherent sigma field state with the property (GHGU Eq. [3.7]): $\bra \Sigma | \hat\sigma(r)|\Sigma\ket = \sigma(r)$, and $|P^0_0\ket$ ($|P^{1 w}_{1  m}\ket$) are pion coherent-pair states to be determined. The normalization of the nucleon state requires $\alpha^2+\beta^2+\gamma^2=1$. 
The permutation symmetric form of the $SU(2)\times SU(2)\times SU(2)$ quark wavefunctions imply that the source terms in the pion field equations will induce  an angular momentum-isospin correlation for the pion field.  Thus, since the pion is isovector, the only allowed angular momentum of the pion will be $\ell=1$, so in the treatment to follow only the $\ell =1$ term of the pion field expressions Eqs.[\ref{pionfield}] and [\ref{pionmom}] is retained.

One constructs a pionic coherent-pair state with quantum numbers of the vacuum as follows \cite{bolsterli}.  Consider the scalar-isoscalar coherent state,
\begin{equation}
|P^0_0\ket=\sum_{n}\frac{f_{n}}{(2n)!}[\hat b^{ 1~\dagger}_1:\hat b^{1~\dagger}_1]^{n} |0\ket,  \label{preln}
\end{equation}
where the double-dot notation refers to spin-isospin (i.e. \{m,w\}) scalar contractions.  A coherent state with spin-isospin=\{11\}, $|P^{1 w}_{1 m} \ket$,  is constructed by operation of 
$(-)^{m+w}\ \hat{b}^{1-w}_{1 -m}$   upon  $|P^0_0\ket$.  Sequential such operations appropriately contracted 
are assumed to close, yielding a recurrence relation for $f_n$, namely,
\begin{equation}
(-)^{m+w}\ \hat{b}^{1 -w}_{1 -m}|P^{00}\ket=a|P^{1 w}_{1 m}\ket.
\end{equation}
A contracted second application gives
\begin{equation}
\sum_{m,w}\hat{b}^{1 w}_{1 m}|P^{1 w}_{1 m}\ket=\tilde{L} b |P^0_0 \ket,
\end{equation}
where the spin-isospin multiplicity is $\tilde L=(2L+1)(2T+1) = 9$ and where closure is forced by associating the vacuum quantum number multipion state uniquely with$ |P^0_0\ket$.  The coherence is determined by the parameter, $x$, defined by
\begin{equation}
(\hat{b}^{1~\dagger}_1:\hat{b}^{1~\dagger}_1)|P\ket=x |P\ket \label{xreln}.
\end{equation}
Here $|P\ket$ can be either $|P^0_0\ket$ or $|P^1_1\ket$ and $x=\tilde{L} a b= 9 ab$ serves as a (as yet free) coherence parameter and the symbol $b^{1~\dagger}_1:b^{1~\dagger}_1$ indicates the coupling to a scalar-isoscalar.  Inserting these into Eq.(\ref{preln}) and Eq.(\ref{xreln}), we obtain the recursion relation (GHGU Eq. [4.11]),
\begin{equation}
f_{n+1}=\frac{x(2n+1)}{(\tilde{L}+2n)}f_{n},
\end{equation}
which can be solved to give (GHGU Eq. [4.12])
\begin{eqnarray}
f_{n}=\frac{x^{n}(2n-1)!! (\tilde{L}-2)!!}{(\tilde{L}-2+2n)!!} f_0,
\end{eqnarray}
where $f_{0}$ is given by the normalization of $|P^0_0\ket$.  The $a$ and $b$ parameters can be expressed as functions of $x$ from the normalization of $|P^1_1\ket$.  We obtained values for the function $f_{0}$, $a(x)$ and $c(x)$ ($b$ is determined from $a$) different from those of GHGU [1988].  We find  
\begin{equation}
a(x)=\frac{1}{3}\Bigl[\frac{(105+45x^{2}+x^{4})\sinh x -(105+10 x^{2})x \cosh x}{-(15+6x^{2}) \sinh x+(15+x^{2})x \cosh x}\Bigr]^{\frac{1}{2}},
\end{equation}
\begin{equation}
c(x)=\frac{1}{3}\Bigl[1+\frac{-(945+420x^{2}+15x^{4})\sinh x+(945+105x^{2}+x^{4})x \cosh x}{(105+45x^{2}+x^{4})\sinh x-(105+10x^{2})x \cosh x}\Bigr]^{\frac{1}{2}},
\end{equation}
and
\begin{equation}
f_{0}^{-2}=(\tilde{L}-2)!! 2^{\bar{x}-1} \partial_{y}^{\bar{x}-1} \cosh (x),
\end{equation}
with $y=x^2$ and
\begin{equation}
\bar{x}=\frac{(2\tilde L+1)}{2} , \hspace{.2in}\partial _{y}=\frac{\partial}{\partial y}=(\frac{1}{2x})\frac{\partial}{\partial x},
\end{equation}
which should be contrasted with GHGU Eqs. [4.13], [4.21] and [4.22].  In addition we found that we needed another factor for the $|P^1_1\ket$ matrix element of the four-pion term (implied summation over repeated indices),
\begin{equation}
\bra P^1_1|\hat b_\alpha^\dagger \hat b_\beta^\dagger \hat b_\beta \hat b_\alpha|P^1_1\ket= 81 d(x)^2,
\end{equation}
where greek subscripts include both spin and isospin and $d(x)$ is given by
\begin{equation}
d(x)=\frac{1}{9}\Bigl[ \frac{(7560+3465x^2+165 x^4 +x^6) \sinh x-(7560+945x^2+18x^4) x  \cosh x}
                                              {(105+45x^2+x^4)\sinh x-(105+10x^2)x \cosh x}\Bigr]^{1/2}.
\end{equation}
 
\section{The Field Equations}

The total energy of the baryon is given by
\begin{equation}
E_{B}=<BT_{3}J_{z}|\int_{0}^{\infty} d^{3}r:\hat{H}(r):|BT_{3}J_{z}>,
\end{equation}
where $B =  N$ or $\Delta$. The field equations are obtained by minimizing the total energy of the baryon with respect to variations of the fields, $\{ u(r),w(r),\sigma(r),\phi(r)\}$, as well as the Fock-space parameters, $\{\alpha,\beta,\gamma\}$ subject to the normalization conditions.  The total energy of the system is written as
\begin{equation}
E_{B}=4\pi\int_{0}^{\infty} drr^{2} {\cal E}_{B}(r).
\end{equation}
We find the following result for the energy density which differs from GHGU Eq. [5.3].  The differences can be traced to different results for coherent-pair matrix elements.  Writing the quark Dirac spinor as
\begin{equation}
\psi^{\frac{1}{2} w}_{\frac{1}{2} m}(\vec r) = \left( \begin{array}{c}u(r)\\ v(r) \vec\sigma\cdot\hat r \end{array}\right)\chi_{\frac{1}{2} m}\xi^{\frac{1}{2} w},
\end{equation}
the energy density is given by
\begin{equation}
\begin{array}{ll}
{\cal E}_{B}(r) = &\frac{1}{2}(\frac{d\sigma}{dr})^{2}+\frac{\lambda^{2}}{4}(\sigma^{2}(r)-\nu^{2})^{2}-m^{2}_{\pi}f_{\pi}\sigma(r)\\
&+3[u(r)(\frac{dv}{dr}+\frac{2}{r}v(r))-v(r)\frac{du}{dr}+g\sigma(r)(u^{2}(r)-v^{2}(r))]\\
& +(N_{\pi}+x)((\frac{d\phi}{dr})^{2}+\frac{2}{r^{2}}\phi^{2}(r))+(N_{\pi}-x)\phi^{2}_{p}(r)\\
& -\alpha\delta (a+b) u(r)v(r)\phi(r)\\
&+\lambda^{2}[x^{2}+2xN_{\pi}+81(\alpha^{2}a^{2}c^{2}+(\beta^{2}+\gamma^{2})d^{2}]\phi^{4}(r)\\
&+\lambda^{2}(N_{\pi}+x)(\sigma^{2}(r)-\nu^{2})\phi^{2}(r),
\end{array}
\end{equation}
where  $N_{\pi}$ is the average pion number (GHGU Eq. [5.12]),
\begin{equation}
N_{\pi}=9[\alpha^{2}a^{2}+(\beta^{2}+\gamma^{2})c^{2}],
\end{equation}
and where $\delta$ takes the following values for nucleon or delta quantum numbers,
\begin{eqnarray}
\delta_N=(5\beta+4\sqrt{2}\gamma)/\sqrt{3} \\ 
\delta_\Delta=(2\sqrt{2}\beta+5\gamma)/\sqrt{3}. 
\end{eqnarray}

The function $\phi_{p}(r)$ is obtained from $\phi(r)$ by the double folding,
\begin{equation}
\phi_{p}(r)=\int_{0}^{\infty}\omega(r,r^{\prime})\phi(r^{\prime})r^{\prime 2} dr^{\prime},
\end{equation}
with
\begin{equation}
\omega(r,r^{\prime})=\frac{2}{\pi}\int_{0}^{\infty} dk k^{2}\omega(k)j_{1}(kr)j_{1}(kr^{\prime}).
\end{equation}
For fixed $\alpha,\beta,$ and $\gamma$ the stationary functional variations are expressed by 
\begin{equation}
\delta[\int_{0}^{\infty} drr^{2}({\cal E}_{B}(r)-3\varepsilon(u^{2}(r)+v^{2}(r))-2\kappa\phi\phi_{p}(r))]=0,
\end{equation}
where the Lagrangian parameter, $\kappa$,  enforces the pion normalization condition,
\begin{equation}
8\pi\int_{0}^{\infty} \phi(r)\phi_{p}(r)r^{2}dr=1,
\end{equation}
and the Lagrangian parameter $\varepsilon$ fixes the quark normalization,
\begin{equation}
4\pi\int_{0}^{\infty} drr^{2}(u^{2}(r)+v^{2}(r))=1.
\end{equation}

Minimizing the Hamiltonian yields the four nonlinear coupled differential equations,
\begin{equation}
\begin{array}{l}
\frac{du}{dr}=-(g\sigma+\varepsilon)v(r)-\frac{2}{3}\alpha\delta(a+b)g\phi(r) u(r),\\
\frac{dv}{dr}=-\frac{2}{r}v(r)-(g\sigma(r)-\varepsilon)u(r)+\frac{2}{3}\alpha\delta(a+b)g\phi(r) v(r),\\
\frac{d^{2}\sigma}{dr^{2}}=-\frac{2}{r}\frac{d\sigma}{dr}-m_{\pi}^{2}f_{\pi}+3g(u^{2}(r)-v^{2}(r))+2\lambda^{2}(N_{\pi}+x)\phi^{2}(r)\sigma(r)  +\lambda^{2}(\sigma^2(r)-\nu^2)\sigma(r), \\
\frac{d^{2}\phi}{dr^{2}}=-\frac{2}{r}\frac{d\phi}{dr}+\frac{2}{r^{2}}\phi(r) +\frac{1}{2}(1-\frac{x}{N_{\pi}})m^{2}_{\pi}\phi(r)\\
\hspace{.4in}+\frac{\lambda^{2}}{2}(1+\frac{x}{N_{\pi}})(\sigma^{2}(r)-\nu^{2})\phi(r)\\
\hspace{.4in}+\frac{\lambda^{2}}{N_{\pi}}(x^{2}+2N_{\pi}x+81(\alpha^{2}a^{2}c^{2}+(\beta^{2}+ \gamma^{2})d^{2})]\phi^{3}(r)\\
\hspace{.4in}-\frac{\alpha}{N_{\pi}}(a+b)g \delta  u(r) v(r)-\frac{\kappa}{N_{\pi}}\phi_{p}(r),
\end{array}
\end{equation}   
with eigenvalues $\varepsilon$ and $\kappa$ .  These consist of two quark equations for $u$ and $v$ where $\sigma(r)$ and $\phi(r)$ appear as potentials, and two Klein-Gordon equations with $u(r)v(r)$ and $(u^{2}(r)-v^{2}(r))$ as source terms.  The boundary conditions are for $r\to 0$,
\begin{equation}
v=\frac{d\sigma}{dr}=\phi= \frac{du}{dr}=0,
\end{equation}
and for $r\to \infty$,
\begin{equation}\begin{array}{l}
\bigl[ r (g f_\pi -\varepsilon)^{1/2}+(g f_\pi +\varepsilon)^{(-1/2)}\bigr] u(r)-r (g f_\pi +\varepsilon)^{1/2} v(r)=0,\\
(2+2 m_\pi r +m^2_\pi r^2)\phi(r) +r(1+m_\pi r)\phi^{\prime}(r)=0,\\
r \sigma^{\prime}(r) +(\sigma(r)-f_\pi)(1+m_\sigma r)=0,
\end{array}
\end{equation}
which has one sign in the first equation different from GHGU Eq. [5.18].
The field equations are solved for fixed coherence parameter, $x$, and fixed Fock-space parameters, $\{\alpha,\beta,\gamma\}$.  Then the expectation value of the energy is minimized with respect to  $\{\alpha,\beta,\gamma\}$ by diagonalization of the ``energy matrix'',
\begin{equation}
  \left[ \begin{array}{ccc} 
      H_{\alpha \alpha} & H_{\alpha \beta} & H_{\alpha \gamma} \\
      H_{\alpha \beta} & H_{\beta \beta} & H_{\beta \gamma} \\
      H_{\alpha \gamma} & H_{\beta \gamma} & H_{\gamma \gamma} \\
  \end{array} \right]
  \left[  \begin{array}{c} \alpha \\ \beta \\ \gamma \end{array} \right]
  = E \left[  \begin{array}{c} \alpha \\ \beta \\ \gamma \end{array} \right].
\end{equation}
Each $H$ entry of the matrix is related to a corresponding density, $E(r)$,  as 
follows:
\begin{equation}
 H_{\alpha \beta} = 4 \pi \int_0 ^{\infty} r^2 E_{\alpha \beta} (r) \ dr, 
\end{equation}
and analogously for the other entries.  The $E_{\alpha\beta}(r)$ functions for a nucleon are
\begin{eqnarray}
 E_{\alpha \alpha}(r) &=& E_0(r) + 18 a^2 \phi_p^2(r) + 9 a^2 \lambda^2(2 x + 9 c^2) \phi^4(r)     \\
   & & + 9 a^2 \lambda^2[\sigma^2(r) - \nu^2] \phi^2(r), \nonumber  \\
 E_{\beta \beta}(r) &=& E_0(r) + 18 c^2 \phi_p^2(r) + 9 \lambda^2(2 x c^2  + 9 d^2) \phi^4(r)     \\
   & & + 9 c^2 \lambda^2 [\sigma^2(r) - \nu^2] \phi^2(r),  \nonumber \\
 E_{\alpha \beta}(r) &=& -2 g(a+b) \phi(r) u(r) v(r) \cdot \frac{5}{\sqrt{3}},  \\
 E_{\alpha \gamma}(r) &=&  -2g (a+b)\phi(r) u(r) v(r)  \cdot 4\sqrt{\frac{2}{3}},  \\
 E_{\gamma \gamma}(r) &=& E_{\beta \beta}(r), \\
 E_{\beta \gamma}(r) &=& 0. 
\end{eqnarray}
The $E_{\alpha\beta}(r)$ functions for a delta are the same except for
\begin{eqnarray}
 E_{\alpha \beta}(r) &=& -2 g(a+b) \phi(r) u(r) v(r) \cdot \frac{2\sqrt{2}}{\sqrt{3}}, \nonumber \\
 E_{\alpha \gamma}(r) &=&  -2g (a+b)\phi(r) u(r) v(r)  \cdot \frac{5}{\sqrt{3}},  \nonumber 
\end{eqnarray}
which expresses the difference in the $\delta$-terms (Eqs. [32-33]) appropriate to nucleon and delta respectively.

In the above expressions, $E_0(r)$ is given by
\begin{eqnarray}
  E_0(r) &=& \frac{1}{2} \left[ \frac{d \sigma}{dr} \right]^2 + 
   \frac{\lambda^2}{4} [\sigma^2(r) - \nu^2]^2 - m_{\pi}^2 f_{\pi} \sigma(r) \nonumber \\  
  & &  + U_0 + \lambda^2 x^2 \phi^4(r) + 3 g \sigma(r)[u^2(r) - v^2(r)]  \nonumber\\
  & &  +3[u(r)(\frac{dv}{dr}+\frac{2}{r} v(r))-v(r)\frac{du}{dr}]-x m_{\pi}^2 \phi^2(r)\nonumber \\
  & &  + \lambda^2 x [\sigma^2(r) - \nu^2] \phi^2(r). 
\end{eqnarray}
Substituting for $U_0$ and using the field equations, this can be rewritten as
\begin{eqnarray}
E_0(r)     & = & \frac{1}{2} [\sigma'(r)]^2 + \lambda^2 x^2 \phi(r)^4 + 3 g \sigma(r)
   (u(r)^2 - v(r)^2) + \frac{\lambda^2}{4}(\sigma(r)^2 - f_{\pi}^2)^2 + 
   \frac{m_{\pi}^2}{2}(\sigma(r)^2 - f_{\pi}^2) \nonumber \\
&& - m_{\pi}^2 f_{\pi}  (\sigma(r) - f_{\pi})+ \lambda^2 x (\sigma(r)^2 - f_{\pi}^2) \phi(r)^2 -2  x m_{\pi}^2\phi(r)^2, 
\end{eqnarray}
which shows an explicit rapid decay as $r\to \infty$. 

We solve this set of equations in the same iterative manner as GHGU.  The iteration procedure is implemented as follows.  For fixed values of $x$ and $\alpha,\beta,\gamma$ the above differential equations with the corresponding boundary conditions are solved by using the same numerical package (COLSYS \cite{colsys}) as used in the original GHGU paper.  Then the energy matrix is diagonalized and the minimum eigenvector  chosen.  These solutions are then mixed with the previous solution and repeated until self-consistency is achieved. The procedure is started with an initial guess. We used
the so-called ``chiral circle'' meson field forms \cite{bb}, but the actual starting point does not matter, provided the iterations converge.

\section{Nucleon Properties}
 
In this section we review the several nucleon observables which will be calculated from the solutions arising from the procedure described in the previous section, noting differences with GHGU where they occur.  From the electromagnetic current operator,
\begin{equation}
\hat{J}^\mu_{\rm em} =\overline{\hat\psi} (\frac{1}{6}+\frac{1}{2}\tau_3 )\hat\psi+
\epsilon_{3 \alpha\beta}\hat\phi_\alpha\partial^\mu\hat\phi_\beta,
\end{equation}
one derives the charge and  magnetic moment densities for the neutron and proton,
\begin{eqnarray}
\frac{\rho_{\rm p}(r)}{4\pi e} &=&\alpha^2(u^2+v^2)+\beta^2[\frac{1}{3}(u^2+v^2) +\frac{4}{3}\phi\phi_p]\nonumber\\
&&+\gamma^2[\frac{4}{3}(u^2+v^2)-\frac{2}{3}\phi\phi_p],\\
\frac{\rho_{\rm n}(r) }{4\pi e}&=&\beta^2 [\frac{2}{3}(u^2+v^2)-\frac{4}{3}\phi\phi_p]\nonumber \\
&&+\gamma^2[-\frac{1}{3}(u^2+v^2)+\frac{2}{3}\phi\phi_p],\\
\frac{\mu_{\rm p}(r) }{4\pi e}&=& \frac{r u v}{81}(54 \alpha^2+2\beta^2+\gamma^2+32\sqrt{2}\beta\gamma)\nonumber \\
&&+\frac{x}{729 a^2}(9 a^2+x)(4\beta^2+\gamma^2)\phi^2,\\
\frac{\mu_{\rm n}(r)}{4\pi e} &=&\frac{r u v}{81}(-36\alpha^2-8\beta^2+\frac{1}{2} \gamma^2-32\sqrt{2}\beta\gamma)\nonumber \\
&&-\frac{x}{729 a^2}(9 a^2+x)(4\beta^2+\gamma^2)\phi^2, 
\end{eqnarray}
which differ from GHGU Eq. [6.6] and [6.7] for the magnetic moment densities.  The axial-vector to  vector coupling ratio is given by (GHGU Eq. [6.9])
\begin{equation}
\frac{g_{\rm A}}{g_{\rm V}} = 2\bra N J_z=\frac{1}{2} T_3 =\frac{1}{2} |\int d^3r : [\frac{1}{2} \overline {\hat\psi}\gamma^5 \gamma_3 \tau_0\hat\psi +\hat\sigma\partial^z\hat\sigma\hat\pi_0-\hat\phi_0\partial^z\hat\sigma]: | N J_z=\frac{1}{2} T_3 =\frac{1}{2}\ket,
\end{equation}
from which we find,
\begin{equation}
\frac{g_{\rm A}}{g_{\rm V}} = 4\pi \int_{0}^{\infty} dr r^2\Bigl[ \bigl[\frac{5}{3} \alpha^2+\frac{5}{27}\beta^2+\frac{25}{27}\gamma^2+\frac{32\sqrt{2}}{27}\beta\gamma\bigr](u^2-v^2/3)+\frac{8}{3\sqrt{3}}\alpha\beta(a+b)\frac{d\sigma}{dr}\phi\Bigr],
\end{equation}
which differs with GHGU Eq. [6.10] in the first term. Finally, the $\pi$NN coupling constant can be calculated from the pion field or the pion source term. Using the pion field form, one has
\begin{equation}
\frac{g_{\pi {\rm NN}}}{2M_N}=4 \pi m_\pi^2\frac{2}{3\sqrt{3}}\alpha\beta(a+b)\int_{0}^{\infty} dr r^3 \phi(r),
\end{equation}
which is in agreement with GHGU Eq. [6.15].  Using the pion source term, one can obtain an  alternative form of the $\pi$NN coupling constant,
\begin{eqnarray}
\frac{g_{\pi {\rm NN}}}{2M_N}&=& 4\pi \int_0^\infty dr r^3 g(\frac{10}{9}\alpha^2+\frac{10}{81}\beta^2+\frac{50}{81}\gamma^2+\frac{64\sqrt{2}}{81}\beta\gamma) u(r) v(r)\nonumber \\ 
&&-4\pi \lambda^2\alpha\beta \int_0^\infty dr r^3 \left[\frac{2}{3\sqrt{3}}(a+b) (\sigma^2(r)-f_\pi^2)\phi(r)+\frac{4}{\sqrt{3}}(a^2 b +2 b^2 a+a c^2)\phi^3(r)\right],
\end{eqnarray}
which differs from GHGU Eq.[6.14] by a factor of 3 in the last term.

The $\sigma$-term was not calculated in GHGU, but is an important quantity in that it measures the degree of chiral symmetry breaking.  For the linear sigma model considered here this quantity is given by
\begin{equation}
\sigma_{\pi N} = 4\pi f_\pi m_\pi^2 \int_{0}^{\infty} dr r^2 (\sigma(r)-f_\pi).
\end{equation}
For a review of this quantity see Ref. \cite{reya}. 

\section{Numerical Results}

As mentioned earlier, the chiral quark model of the nucleon has two free parameters once the pion mass and pion decay constant are fixed at their experimental values.  One can choose the free parameters to be the $\sigma$-mass and the meson-nucleon coupling constant, $g$.  For comparison with GHGU and to illustrate the systematics of the model, for the results to follow, we fix these two parameters to $m_\sigma$= 700 MeV and $g$=5.00.  

First consider the coherence parameter.  Figure 1 shows the baryon (nucleon and delta) energy as the coherence parameter is varied.  As was similarly shown in GHGU's Fig. [1], there is a clear minimum in the region of $x$=1 with a corresponding $N_{\pi}=.43$.  Table I shows the various energy contributions to the baryon mass.  While Table II gives the values of the derived parameters determined by the minimization procedure.  The values found are in good agreement with that found in GHGU (self-consistent case) even though a different coherence dependence (which shows up in the $d$-term term in Eq. [29]) was used.  This is expected however since the fourth-power pion self-interaction energy makes a relatively small contribution to the total energy.  The nucleon-delta mass difference is found to be about 150 MeV. Figure 2 shows the quark wave functions and the meson fields for the $x$=1 case, and Figure 3 shows the pion field and its derived kinetic energy density function, $\phi_p(r)$.  All of these quantities are little changed from those reported by GHGU.

Next consider the nucleon physical properties.  Table II shows the results for several nucleon observables compared with the projected Hedgehog model of Birse \cite{birse} and with experiment.  For those nucleon quantities calculated by GHGU our results are nearly identical despite the differences in several quantities noted previously.  

In comparing the predictions of the coherent-pair model to those of the Hedgehog model, we find that the quark contributions to each observable are roughly similar; however the pion contributions are significantly smaller, a feature consistently seen throughout.  This is somewhat surprising given the fact that both approaches are attempting to solve the nucleon problem from the same starting model albeit with very different methods and resulting different model parameters.  Quantities such as the sigma-commutator which depend on the sigma field are quite similar to that of the Hedgehog model, both giving a sigma commutator of roughly 90 MeV.  The small pion contributions to the calculated observables seems to be the coherent-pair approximation's principal phenomenological short-coming.  

Consider the nucleon charge radius squared. For the Hedgehog model the pion's contribution to the proton's charge radius squared  is roughly 40\% 
that of the quarks; while for the coherent-pair approximation the pions contribution is only  4\% 
that of the quarks.  The small pionic contribution is compensated for by a slightly larger quark contribution leaving the total proton charge radius squared very close to that calculated in the Hedgehog model, but still about 20\%
that of the experimental value.  

 Likewise for the magnetic moments.  The quark contribution to the magnetic moments is about the same for both models, but for the Hedgehog model the pions make a contribution to the proton magnetic moment which is 65\%
 that of the quarks while for the coherent-pair approximation the pionic contribution is only 12\% 
that of the quarks. The smaller pionic contribution to the magnetic moments results in magnetic moments roughly 60\% that of the empirical values. 

 For the case of the axial vector coupling constant we find again that the quark contributions in the two models are similar, but the mesonic contribution in the coherent-pair model is only half that of the Hedgehog model. In this case this actually helps in that the Hedgehog model predicts too large a value for this quantity. 

Finally we consider the $\pi$NN coupling constant calculated in two ways first using the pion field itself Eq.[59] and secondly using the pion source term Eq.[60].  In the Hedgehog model Birse found that both methods gave roughly the same value, but in the coherent-pair approximation the two methods give very different results.  This large difference in the two, supposedly equivalent, ways of calculating $g_{\pi NN}$ has been noted already in Fiolhais, et al. \cite{fiolhais} who studied the generalized projected Hedgehog model and compared with other models including GHGU.  In Fiolhais et al. the difference in $g_{\pi NN}$ calculated from the pion field and the pion source terms along with the value expected from the Goldberger-Treiman relation provide a virial measure of how well the self-consistency condition in the pion sector is met.   The Fiolhais et al. relation for the fractional virial deviation is
\begin{equation}
\Delta =\frac{g_{\pi NN}-g'_{\pi NN}}{g^{av}_{\pi NN}} 
          +2\frac{(1.08~g^{GTR}_{\pi NN}-g^{av}_{\pi NN}}{(1.08) g^{GTR}_{\pi NN}+g^{av}_{\pi NN}},
 \end{equation}
where $g_{\pi NN}$ is the source-calculated value, Eq. [60], $g'_{\pi NN}$ is field-calculated value, Eq. [59], $g^{av}_{\pi NN}$ is the average of these two, and $g^{GTR}_{\pi NN}$ is the value expected from the Goldberger-Treiman relation,
\begin{equation}
g^{GTR}_{\pi NN}=\frac{M_N}{f_{\pi}}\frac{g_A}{g_V}.
\end{equation} 
The factor of 1.08 accounts for the effects of explicit chiral symmetry breaking. 
Using the coherent-pair approximation results of GHGU, Table 3 of Fiolhais et al. reports a fractional virial violation of 173\%.  This should be compared with 51\% for the mean-field Hedgehog model of Birse and Bannerjee and 6\% for the generalized projected Hedgehog model of Fiolhais et al.  Using the values in Table III, we find a fractional virial violation of 149\% which is a little better than that previously reported, but still shows a clear and substantial problem with self-consistency in the pion sector.   

    One must question why this should be so.  That the two models should give such different pionic contributions is a bit puzzling as the starting model is the same and both employ a self-consistent mean-field type  approach which minimizes the energy.  In comparing the quark wavefunctions and meson field solutions in the two models one finds, not surprisingly, that the quark wavefunctions and sigma field are nearly identical in the two models, but, though  the pion field function in the coherent-pair approximation is smaller by a factor of about two from  that found in the Hedgehog model, this cannot be the source of the discrepancy since the pion field in the coherent-pair approximation is normalized but the magnitude of the pion field (treated as a classical field)  in the Hedgehog model is determined by the source terms.  The magnitude of the pionic contribution to any quantity in the coherent-pair approximation is determined by the Fock space coefficients, $\{ \alpha, \beta, \gamma \}$ and by the various coherent functions $\{ a, b, c, d \}$.  With the minimized solution around $x\sim 1.0$ the mean number of pions in the coherent pair approximation case is $N_\pi = 0.43$ which apparently yields the small pionic contributions to the various baryon observables.  A comparison of the magnitude of the pion field resulting from the Hedgehog model with the normalized field arising in the coherent-pair model would imply an effective number of pions about 3 times greater in the Hedgehog model.  A preliminary search of the parameter space to see if some other parameter set may correct this deficiency was not successful.  We have examined the model up to a coherence parameter of $x$=2.25.  Larger coherence parameters do indeed give rise to greater pionic contributions to the various nucleon observables, but at the expense of greater energy as well.  While the issue is being investigated further, we suspect the problem with self-consistency as revealed in the large fractional virial violation to be the principal difficulty.

Another useful comparison is shown in Figs. \ref{kineticfig} and \ref{qmesonfig} where the energy densities for various terms are shown for both the Hedgehog model and the coherent-pair approximation.  Again there is a systematically smaller contribution to each term in the coherent-pair model than in the Hedgehog model.  Yet the total energy of both solutions are roughly the same (1070 MeV and 1120 MeV respectively).

\section{Summary and Conclusions}

In this work we re-examined a linear sigma model of the nucleon using quarks and sigma and pion mesons as the fundamental degrees of freedom.  To solve this model we have employed the coherent-pair approximation following the work of Goeke, et al. correcting several errors.  We solved the model  using a Fock space ansatz treating the sigma field as a classical field and treating the pions as quantum fields using the coherent-pair approximation of Bolsterli \cite{bolsterli}.  We neglect vacuum effects and center-of-mass corrections. Despite the several corrections to the work of Goeke, et al., the numerical solutions we find are very close to those presented in their original paper  as are all the calculated  nucleon observables.  We find that the calculated nucleon observables are reasonably close to experiment, but, in the case of electromagnetic quantities such as charge radii or magnetic moments, the pionic contributions seem too small when compared to that of other chiral nucleon models such as the Hedgehog model of Birse and Bannerjee \cite{{bb},{birse}}.  The origin of this difference is not fully understood but probably arises from the lack of self-consistency in pionic sector using this approach as was noted previously by Fiolhais, et al..  Therefore, at this stage we must concur with the rather disappointing conclusion of Goeke, et al., namely, that a better description of the pionic sector is required before a more satisfactory nucleon phenomenology can be expected in this model.

%
%
%
\section{Acknowledgements}
We thank J. R. Shepard and V. Dmitrasin\'ovic for helpful comments and suggestions. We also thank K. Goeke for helpful remarks and for providing additional useful references. T. S. T. Aly thanks the government of Egypt for supporting his graduate work at the Colorado School of Mines.

\newpage
%
%
%

%

%
%
%

\begin{table}
\caption{ Energy contributions to coherent-pair nucleon using g=5 and 
$m_{\sigma}$=700 MeV. (All values are in MeV.)  Self-consistent solution evaluated using a coherent parameter value $x$=1.  The details of this solution are discussed in the text.}
\begin{tabular}{lrr}
Quantity & Nucleon&  Delta \\
\tableline
Quark eigenenergy                 & 150 &  219\\
Quark kinetic energy              & 1124  &975\\
Sigma kinetic energy               &  304 &  268\\
Pion kinetic energy               &   236 &185\\
Quark-meson interaction energy &  --675 & --318\\
Meson interaction energy          &   84&114\\
Baryon Mass                       &  1073 &1224\\
Nucleon-Delta mass difference&& 140
\end{tabular}
\end{table}

\begin{table}
\caption{ Numerical values of various parameters for the coherent-pair nucleon with a coherence parameter of $x$=1 and using $g$=5 and 
$m_{\sigma}$=700 MeV after self-consistent minimization.  (N/A means ``not available''.) }
\begin{tabular}{lrr}
Quantity & This work& Goeke, et al.\cite{goeke} \\
\tableline
Mass (GeV)                 & 1.073 &  1.08\\
$\alpha$              &  .820&  .82\\
$\beta$           &   .379 &.38\\
$\gamma$  &  .429& .43\\
Pion eigenvalue, $\kappa$ (GeV)   &   --.173& N/A\\
Quark eigenvalue, $\epsilon$ (GeV)      &  .150&.14\\
$N_\pi$      &  .43&N/A
\end{tabular}
\end{table}

\begin{table}
\caption{Observables for the coherent-pair nucleon with a coherence parameter of $x$=1 and  using $m_\sigma$=700 MeV and $g$ = 5.  Magnetic moments are in nuclear magnetons. Charge radius is in fm. For comparison the results from the projected Hedgehog model of Birse [5] is also presented.} 
\begin{tabular}{cccccccc}
  &&Coherent Pair &&&  Hedgehog  \cite{birse}\\
Quantity&Quark & Meson& Total&Quark&Meson&Total&Exp.\\
\tableline
$\langle r^2\rangle_{ch-proton}$ (fm$^2$)& 0.533&.023&0.556&0.39&0.16&0.55&0.70\\
$\langle r^2\rangle_{ch-neutron}$ (fm$^2$)& 0.019&--.023&--0.004&0.09&--0.16&--0.070&--0.12\\
$\langle \mu\rangle_{proton}$ &1.53 &.18&1.71&1.74 & 1.13&2.88 &2.79\\                   
$\langle \mu\rangle_{neutron}$ &--1.13&--.18&--1.31&--1.16&--1.13&--2.29&--1.91\\                 
$\frac{g_A}{g_V}$&1.07&.39&1.46&1.11&.75&1.86&1.25\\
$g_{\pi NN}{m_{\pi}\over 2 M}$(Eq.[59])&&&0.25&&&0.93&1.0\\
$g_{\pi NN}{m_{\pi}\over 2 M}$(Eq.[60])&1.11&.24&1.35&1.16&.379&1.53&1.0\\
Sigma term (MeV)&&&88.9&&&94.&$\sim$45 $\pm$5 \cite{reya}
\end{tabular}
\end{table}

%
\begin{figure}
\centerline{\psfig{figure=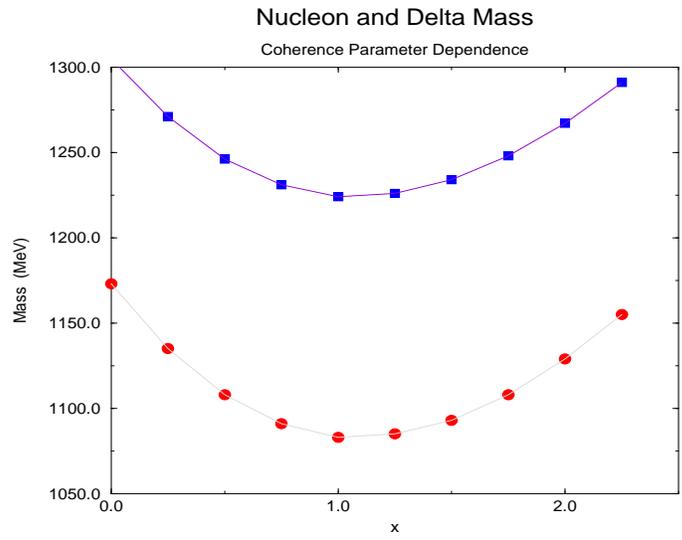,height=3in,width=4in,angle=-90}}
\caption{Dependence of chiral nucleon and delta mass with respect to the coherence parameter, $x$, using
$g$=5 and $m_\sigma=$700 MeV.}
\label{xfig}
\end{figure}
\vskip 1in

\begin{figure}
\centerline{\psfig{figure=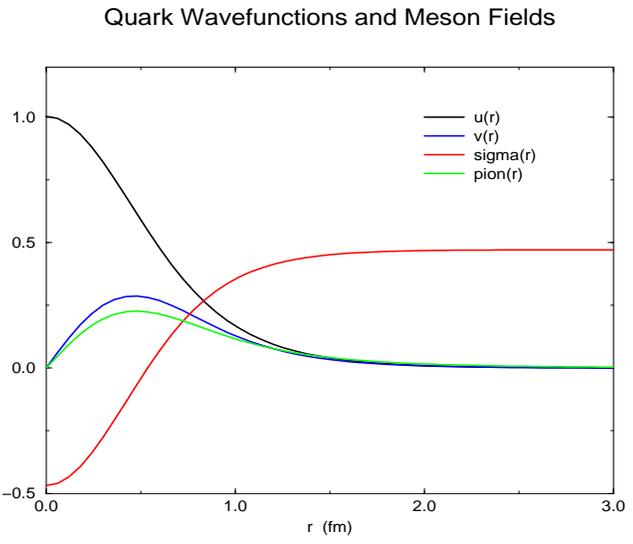,height=3in,width=4in,angle=-90}}
\caption{Quark wavefunctions and meson fields using coherence parameter of $x$=1 with $g$=5 and 
$m_\sigma=$700 MeV.}
\label{fieldsfig}
\end{figure}
\vskip 1in

\begin{figure}
\centerline{\psfig{figure=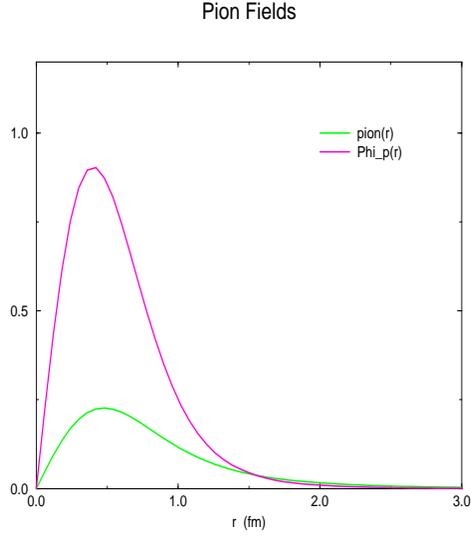,height=3in,width=3in,angle=-90}}
\caption{Pion field shown with kinetic energy pion field function, $\Phi_p(r)$, using coherence parameter of $x$=1 with $g$=5 and $m_\sigma=$700 MeV.}
\label{pionfig}
\end{figure}
\vskip 1in

\begin{figure}
\centerline{\psfig{figure=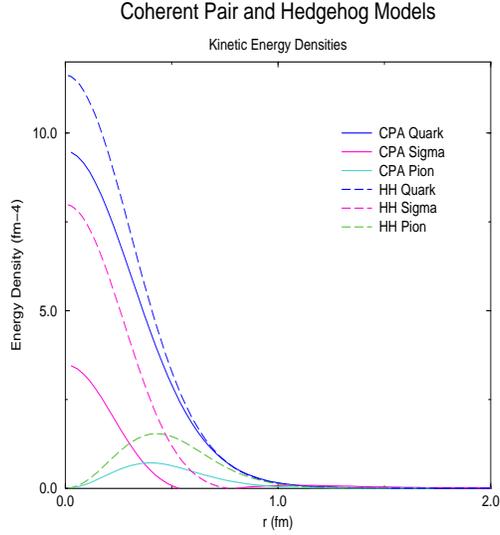,height=3in,width=3in,angle=-90}}
\caption{Comparison of kinetic energy densities for the Hedgehog model  and the coherent-pair approximation  using coherence parameter of $x$=1 with $g$=5 and $m_\sigma=$700 MeV.}
\label{kineticfig}
\end{figure}
\vskip 1in

\begin{figure}
\centerline{\psfig{figure=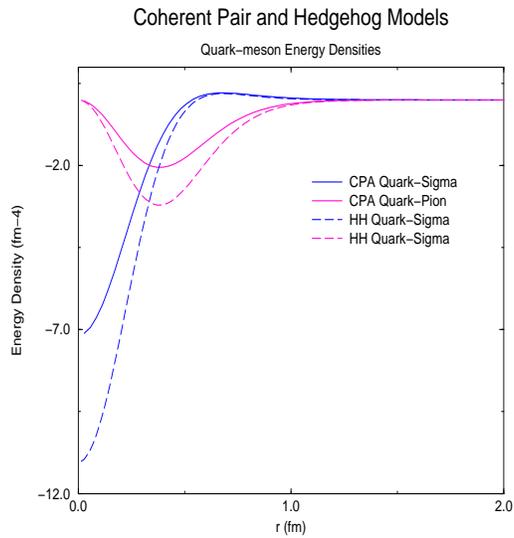,height=3in,width=3in,angle=-90}}
\caption{Comparison of quark-meson interaction energy densities for the Hedgehog model and the coherent-pair approximation  using coherence parameter of $x$=1 with $g$=5 and $m_\sigma=$700 MeV.}
\label{qmesonfig}
\end{figure}
\vskip 1in

\end{document}